\documentclass{article}
\usepackage{spconf,amsmath,graphicx,amsfonts,amssymb,booktabs}

\title{A Signal Subspace Rotation Method for Localization of Multiple Wideband Sound Sources}

\name{Kainan Chen,$^{1}$
      Wenyu Jin,$^{1,2}$
      Bharadwaj Desikan,$^{1}$
      }
\address{$^1$ Huawei German Research Center, Munich, Germany\\
			\{kainan.chen, bharadwaj.desikan\}@huawei.com\\
         $^2$ Department of Algorithm, \\Starkey Hearing Technologies, Eden Prairie, United States\\
         wenyu.jin@ieee.org\\
}

\begin{document}
%
\maketitle
\begin{abstract}

In this paper, the problem of extending narrowband multichannel sound source localization algorithms to the wideband case is addressed. The DOA estimation of narrowband algorithms is based on the estimate of inter-channel phase differences (IPD) between microphones of the sound sources. A new method for wideband sound source DOA estimation based on signal subspace rotation is present. The proposed algorithm normalizes the narrowband signal statistics by rotating the estimated signal subspace to the wideband counterpart in the eigenvector domain. Then the wideband DOA estimate can be obtained by estimating the normalized IPD from these wideband signal statistics.
In addition to requiring less computational complexity compared to repeating the narrowband algorithms for all relevant frequencies of wideband signals, the proposed method also  does not require any additional prior knowledge. The experimental results demonstrate the efficacy and the robustness of the proposed method.

\end{abstract}
\begin{keywords}
Sound source localization, wideband, signal subspace rotation
\end{keywords}
\section{Introduction}
\label{sec:intro}
Sound source localization is an important component in many multichannel signal processing systems aiming, e.g., at source tracking, signal separation, enhancement and noise suppression~\cite{jin17,jin18} .
Traditional sound source localization algorithms such as GCC-PHAT~\cite{phat} estimates the time delay of arrival (TDOA) between a pair of microphones to localize a single source. SRP-PHAT~\cite{srpphat}  as an extension can further localize multiple sources simultaneously.
Another group of algorithms that resolves the simultaneous multiple source localization problem is based on high-resolution subspace techniques, such as MUSIC~\cite{music} and ESPRIT~\cite{esprit} .
For highly reverberant senarios, the Independent Component Analysis (ICA) based algorithms are proposed ~\cite{lombardtrans,nesta2011}.

ESPRIT~\cite{esprit} exploits the algebraic properties of the spatial covariance matrix. This method features good performance for narrowband signals. However, it is not directly applicable to wideband signals.
Hence, it process the wideband signals by estimating the individual narrowband results in the Short Time Fourier Transform (STFT) domain.
Post-processing schemes such as the histogram method only exploit the narrowband estimation results while the mutual information between different frequency bins is not considered.
Especially for known type of the sound sources, such as speech, the process based on clustering STFT bins such as~\cite{andreas2018cluster,araki2006doa} can further improve the accuracy.


Many studies focus on extensions of narrowband localization algorithms to wideband signals. The methods of \cite{eb_esprit,Khaykin2009Coherent} transform signals to cylindrical/spherical harmonics for circular/spherical microphone arrays. Then the rotational invariances are normalized through frequencies and the further estimation via ESPRIT is based on wideband signals.
A Coherent Signal Subspace (CSS) method that derives the so-called focusing matrices based on steering vectors was introduced in~\cite{css}. The focusing matrices aim to adapt each narrowband signal covariance matrix to generalize to the wideband case. Theoretical analysis and experiments show that the focusing matrices lead to a wideband MUSIC algorithm~\cite{css}, and in~\cite{cssesprit1,cssesprit2,focusingmatrix}, the focusing matrices extend ESPRIT to wideband signals. As pointed out in~\cite{espritextension,tensoresprit}, the focusing matrices method adapts the covariance matrices via linear transformations that are dependent on frequency and DOAs. Therefore, these extensions require a priori knowledge of DOA estimates and on the array manifold for initializing the transformation matrices. To avoid this requirement, the algorithm in~\cite{espritextension} estimates an AR model for the transmission channel of the wideband signals as a priori knowledge for wideband localization. The performance of this algorithm crucially depend on how well the AR model is estimated. However, the model that they proposed and the conventional models are not suitable for non-stationary scenarios.

A novel method that adapts the narrowband ESPRIT to wideband signals is introduced in this work.
The key idea is the rotation of the eigenvectors that span the signal subspace by the corresponding frequency, and to reconstruct the covariance matrices for each frequency bin based on the rotated eigenvectors. In comparison with the focusing matrices approach, the proposed method utilize the wideband signal second-order statistics while does not require a priori knowledge of the DOA which is typically obtained by repeating narrowband MUSIC~\cite{focusingmatrix}.
Experiments with simulated and real recordings show the advantages of the new method regarding performance and computational complexity.

\section{Signal model}
\label{sec:model}

We assume a microphone array which fits the requirements for using ESPRIT, i.e., to localize $Q$ sources.
We use $P$ microphones ($P>Q$) that form a linear array with a uniform spacing $\Delta d$ and with mutually uncorrelated zero-mean sensor noise. The environment is assumed with free-field and far-field conditions. The source-microphone model in the STFT domain is defined as
\begin{equation}
  \label{eq:sourcemodel}
	{X}_p(f_i) = \sum_{q=1}^Q {A}_{p,q}(f_i){S}_q(f_i) + {N}_p(f_i),
\end{equation}
where ${X}_p(f_i)$ and ${N}_p(f_i)$ denote the observation and the noise at the $p$-th microphone in the STFT domain at frequency bin $f_i$, respectively, ${S}_q(f_i)$ denotes the $q$-th source signal, and $A_{p,q}(f_i)$ denotes the frequency response of the propagation path which are given by the rows of ${A(f_i)}$  for $q$-th source arriving at the $p$-th microphone.

The power spectral density matrix for all Q source signal components at the $i$-th frequency bin $f_i$ is denoted as $\mathbf{R}_s(f_i)$ and the power spectral density matrix for the observed noise is denoted as $\mathbf{R}_n(f_i)$. The narrowband component covariance matrix $\mathbf{R}(f_i)$ of the microphone signals can be expressed as
\begin{equation}
  \label{eq:covmodel}
	\mathbf{R}(f_i) = \mathbf{A}(f_i)\mathbf{R}_{s}(f_i)\mathbf{A}^\text{H}(f_i) + \mathbf{R}_{n}(f_i),
\end{equation}
where $*^\text{H}$ denotes the Hermitian transpose, $\mathbf{A}(f_i)$ denotes a $P\times Q$ matrix that captures the steering vectors at the frequency bin $f_i$. The element of $\mathbf{A}(f_i)$ at the $p-{\text{th}}$ row and $q-{\text{th}}$ column, $A(f_i)_{p,q}$, can be expressed as
\begin{align}
	\label{eq:adef}
	\begin{split}
		\begin{gathered}
			A(f_i)_{p,q} = e^{-j2\pi f_i\Delta t_{p,q}}, \\
			\Delta t_{p,q} \stackrel{\text{def}}{=} (p-1)\Delta d c^{-1}\sin{\theta_q}.
		\end{gathered}
	\end{split}
\end{align}
Therefore, the steering vectors are frequency-dependent and the relationship between different frequency bins $f_1,f_2$ can be expressed as,
\begin{align}
	\label{eq:steerrotate}
	\begin{split}
		\begin{gathered}
			\mathbf{A}(f_1)^{\circ f_1^{-1}} = \mathbf{A}(f_2)^{\circ f_2^{-1}},
		\end{gathered}
	\end{split}
\end{align}
where $\lbrace *\rbrace^{\circ f^{-1}} $ describes element-wise exponentiation by $f^{-1} $ and $f_1,f_2$ denote any two frequencies below the spatial aliasing frequency $f_a$,
\begin{align}
    \label{eq:hadamardpower}
	\begin{split}
		\begin{gathered}
			f_a = \lfloor \frac{c}{2\Delta d \sin{\theta}}\rfloor,
		\end{gathered}
	\end{split}
\end{align}
where $\lfloor * \rfloor$ denotes a function that returns the next lower frequency bin.

For our proposed evolution of ESPRIT to wideband source localization, the first step is to estimate the narrowband steering vector. Then the IPDs of each source are obtained from the estimation result. Therefore, the IPDs are depending on frequency and ESPRIT has to be repeated for each narrowband to localize wideband sources, as it is also the case for other narrowband localization algorithms ~\cite{music,broadbeam}. A solution to estimate a unique IPD for each source through the frequency bins is to rotate the estimated steering vector based on (\ref{eq:steerrotate}) as it is shown in the following.

\section{Proposed Approach}
\label{sec:proposed}
The proposed approach considers a novel way for adapting  ESPRIT to a single or multiple wideband sources. In case of a single source scenario, for each narrowband component, the proposed approach rotates the eigenvectors that span the signal subspaces by normalizing the IPD. In case of multiple sources, it reconstructs covariance matrices from the rotated eigenvectors.

To obtain an estimate of the signal subspace, the least-squares (LS) criterion is conventionally employed to find $Q$ vectors that describe the signal subspace and $P-Q$ noise vectors to represent the noise subspace in an LS sense  (see, e.g.,~\cite{music,esprit}). For $Q$ independent target sources, the $P\times P$ matrix  $\mathbf{A}(f_i)\mathbf{R}_s(f_i)\mathbf{A}^\text{H}(f_i)$ is at least of rank $Q$ and positive semidefinite by construction. Therefore, by taking the eigenvalue decomposition of $\mathbf{R}(f_i)$, the eigenvectors corresponding to the largest $Q$ eigenvalues are assumed to be optimum to span the signal subspace at the frequency bin $f_i$ (see, e.g.,~\cite{esprit}).

The eigenvalue decomposition of $\mathbf{R}(f_i)$ can be expressed as
\begin{equation}
  \label{eq:evd}
	\mathbf{R}(f_i)\mathbf{U}(f_i) = \mathbf{U} (f_i)\pmb{\Lambda} (f_i),
\end{equation}
where $\mathbf{U}$ denotes the eigenvector matrix and $\pmb{\Lambda}$ is a diagonal matrix that contains the corresponding eigenvalues. The eigenvector spans the signal subspace is denoted as $\mathbf{U}_s$, and the eigenvector for the noise subspace is denoted as $\mathbf{U}_n$,
\begin{equation}
  \label{eq:eigenvector}
	\mathbf{U}(f_i) = \left[\mathbf{U}_s(f_i) | \mathbf{U}_n(f_i) \right].
\end{equation}

By the relationship defined in~(\ref{eq:steerrotate}), to rotate the estimated subspaces such that they become frequency-independent for all frequency subbands, the estimated signal subspace rotation is defined as
\begin{equation}
  \label{eq:rotateev}
	\mathbf{U}^{'}(f_i) = \mathbf{U}(f_i)^{\circ f_i^{-1}}.
\end{equation}
Then the frequency component of the IPDs are assumed to cancelled.

ESPRIT is based on source subspace analysis~\cite{esprit}. For the single-source ESPRIT, the estimated source subspace is described by the vector $\mathbf{U}^{'}_s(f_i)$. By weighting and summing the rotated eigenvectors that span the source subspace $\mathbf{U}^{'}_s(f_i)$, the estimated wideband source subspace $\mathbf{U}^{'}_{ss}$ can be obtained as
\begin{equation}
  \label{eq:singlesourceadapt}
	\mathbf{U}^{'}_{ss} = \sum_{i} \beta(f_i)\mathbf{U}^{'}_s(f_i),
\end{equation}
where $\beta(f_i)$ denotes a frequency-dependent weighting factor. The eigenvalues of the sources are relevant to the signal power at a certain frequency and can be assumed to reflect the reliability of the signal subspaces estimates. Therefore, the weighting function is chosen as

\begin{equation}
  \label{eq:weighting}
	\beta(f_i) = \text{trace}\{ \pmb{\Lambda}_s(f_i)\} ,
\end{equation}
where $\text{trace}\{ * \}$ denotes a function which returns the trace of the matrix and $\pmb{\Lambda}_s(f_i)$ denotes the eigenvalue matrix of the signal subspace.

Following ESPRIT~\cite{esprit}, the submatrices satisfy the invariance relation
\begin{align}
  \label{eq:signalsubspacephi}
	&\mathbf{U}^{'}_{ss2} = \mathbf{U}^{'}_{ss1} \mathbf{\Phi}, \\
	\text{with} ~~~&\mathbf{\Phi} = \text{diag}\lbrace e^{-j2\pi \Delta t_{1}}, ... ,e^{-j2\pi \Delta t_{P}}\rbrace
\end{align}
where $\mathbf{U}^{'}_{ss1}$ and $\mathbf{U}^{'}_{ss2}$ denote the vectors contain the first and last $P-1$ elements of $\mathbf{U}^{'}_{ss}$, respectively.
Since the combined vector $\mathbf{U}^{'}_{ss}$ is assumed to span the wideband signal subspace $\mathbf{E}_s$, it holds that
\begin{equation}
  \label{eq:signalsubspace}
	\mathbf{E}_s = \mathbf{U}^{'}_{ss} \mathbf{T},
\end{equation}
where $\mathbf{T}$ is a non-singular matrix. Therefore, the subspaces of the two subarrays can be defined as
\begin{align}
    \label{eq:twosubspaces}
	\begin{split}
		\begin{gathered}
			\mathbf{E}_{s1} = \mathbf{U}^{'}_{ss1} \mathbf{T}, \\
			\mathbf{E}_{s2} = \mathbf{U}^{'}_{ss2} \mathbf{T} =  \mathbf{U}^{'}_{ss1} \mathbf{\Phi} \mathbf{T}, \\
			\mathbf{E}_{s2} = \mathbf{E}_{s1}\mathbf{\Psi},
		\end{gathered}
	\end{split}
\end{align}
where $\mathbf{\Psi}=\mathbf{T}\mathbf{\Phi}\mathbf{T}^{-1}$.  $\mathbf{\Psi}$ can then be obtained from (\ref{eq:twosubspaces}) by applying a standard least-squares or total least-squares solver. By realizing that the eigenvalues of $\mathbf{\Psi}$ are the diagonal elements of $\mathbf{\Phi}$ the locations of the sources can be estimated.

For the multi-source ESPRIT, each estimated narrowband source subspace is described by the columns of the matrix $\mathbf{U}^{'}_s(f_i)$. The order of the columns of the matrix is generally not known and may be different for different frequency bins. Therefore, the estimated signal subspaces cannot be combined using (\ref{eq:singlesourceadapt}).

A simpler and more robust solution than the source subspace identification is to reconstruct estimated signal subspaces back to the form of a $P\times P$ covariance matrix like $\mathbf{R}^{'}(f_i)$. This can be achieved by the inverse process of eigenvalue decomposition~(\ref{eq:evd}) using the frequency component cancelled matrix $\mathbf{U}^{'}(f_i)$,
\begin{equation}
  \label{eq:reconstructr}
	\mathbf{R}^{'}(f_i) = \mathbf{U}^{'}(f_i)\pmb{\Lambda} (f_i)\mathbf{U}^{'}(f_i)^{-1}.
\end{equation}

With the same weighting factor definition in (\ref{eq:weighting}), the reconstructed covariance matrices are calculated by
\begin{equation}
  \label{eq:widebandcov}
	\mathbf{R}^{''} = \sum_i \beta(f_i)\mathbf{R}^{'}(f_i).
\end{equation}
Therefore, the eigenvectors that span the estimated signal subspaces are remixed and accumulated through frequencies in $\mathbf{R}^{''}$. By using the conventional narrowband ESPRIT, wideband signal subspace can be separated back and wideband DOAs can then be estimated.

\section{Implementation}
\label{sec:implementation}

In the estimated signal subspace rotation step~(\ref{eq:rotateev}), when $f$ gets larger, a finer quantization is required to limit the effect of quantization errors on the DOA estimation.
An iterative accumulation method is proposed to solve this numerical sensitivity problem. In each iteration, the estimated signal subspace from the $i$-th $(i\in \mathbb{N_+})$ frequency bin is rotated to adapt the next frequency bin ($i+1$) to reconstruct the covariance matrix $\mathbf{R}^{''}_{i+1}$. After weighting and summing to $\mathbf{R}^{''}_{i}$, the new accumulated covariance matrix $\mathbf{R}^{'''}_{i+1}$ is rotated to the next frequency bin ($\mathbf{R}^{''}_{i+1}$) for the next iteration. The rotation step is then as small as the power of $f_{i+2}f_{i+1}^{-1}$. The iteration process can be expressed as

Initialization:
\begin{equation}
  \label{eq:inititeration}
	\mathbf{R}^{''}_{1} = \mathbf{U}_s^{'f_{1}f_{0}^{-1}}(f_0)\pmb{\Lambda}(f_0)\mathbf{U}_s^{'f_{1}f_{0}^{-1}}(f_0)^{-1}
\end{equation}

In each iteration:
\begin{align}
\label{eq:initeration}
\begin{split}
\begin{gathered}
\mathbf{R}^{'''}_{i+1} = \beta(f_i)\mathbf{U}_s^{'}(f_i)\pmb{\Lambda}(f_i)\mathbf{U}_s^{'}(f_i)^{-1} + \mathbf{R}^{''}_{i} \\
\mathbf{R}^{'''}_{i+1} \stackrel{\text{def}}{=} \mathbf{U}_s^{'''}\pmb{\Lambda^{'''}}\mathbf{U}_s^{'''-1} \\
\mathbf{R}^{''}_{i+1} = \mathbf{U}_s^{'''f_{i+2}f_{i+1}^{-1}}\pmb{\Lambda}^{'''}\mathbf{U}_s^{'''f_{i+2}f_{i+1}^{-1}}
\end{gathered}
\end{split}
\end{align}

Because of the spatial aliasing problem, the iteration continues until the lowest aliasing frequency, denoted by $f_{a_0}$,
\begin{equation}
  \label{eq:aliasingfreq}
	f_{a_0} \stackrel{\text{def}}{=} \lfloor\underset{\theta}{\mathrm{argmin}}f_{a}\rfloor = \lfloor\frac{c}{2\Delta d}\rfloor.
\end{equation}
To utilize higher frequencies, the replication method~\cite{replicationphase} can potentially be useful for detecting the aliasing frequency for each source.
Similar to multi-source localization, the DOA can be obtained using the matrix $\mathbf{R}^{''}(f_a)$.


\section{Evaluation}
\label{sec:evaluation}
A set of experiments was performed in order to evaluate the performance of the algorithm using real-world recordings.
The evaluation includes comparisons to the narrowband ESPRIT with the histogram method (hist-ESPRIT)~\cite{esprithistogram} and to the CSS method~\cite{css}.

\subsection{Experimental setup}
\label{secsec:parametrization}
The recordings were captured by a uniform linear array (ULA) with five microphones in a low-reverberation lab ($\text{T60}\approx 0.2s$ of size $9m\times 8m\times 3m$). The microphone model was AKG C562CM. The spacing between microphones $\Delta d$ was $0.044$m. The background noise was white noise, and it was played back via 22 surround speakers to emulate diffuse background noise. The set of sources contains white sources (independent from the background noise) and a set of speech recordings selected from the GRID Corpus~\cite{gridcopus}.
There were two sources located at an angle of $45^\circ$ and at $-45^\circ$ at a distance of $3m$ from the center of the microphone array. In order to evaluate the robustness of the proposed localization method, three test conditions were used: (1) single white source (2) two competing white sources, (3) two competing talkers.
The total length of the recordings is $800$s. The sampling rate was $16000$ samples per second.

The proposed method, hist-ESPRIT~\cite{esprithistogram} and CSS~\cite{css} were processed block-wise with $50\%$ overlaps. The length of the block was $1024$ samples. The frequency band used for these algorithms was up to $3800$Hz, below the aliasing frequency.

\subsection{Results}
\label{secsec:results}
The scenario of the first experiments consisted of a single white noise as the source under $10$dB and $0$dB background diffuse white noise. The source signal was played in an alternating fashion from two speakers located at $45^\circ$ and $-45^\circ$. The result is shown in Table~\ref{table:alterwhite10db}. The performance of the algorithm is evaluated by the mean absolute error (MAE) and the standard deviation of the error (SDE). It can be seen from the results that the proposed algorithm is more accurate and more stable under this strong background noise scenario.
\begin{table}
\centering
\begin{tabular}{ccc|cc}
 \toprule
  & \multicolumn{2}{c|}{SNR=10dB}& \multicolumn{2}{c}{SNR=0dB} \\
 Algorithm & MAE & SDE & MAE & SDE \\
 \hline
 \midrule
hist-ESPRIT & $2.10^\circ$ & $2.47^\circ$  & $3.44^\circ$ & $2.47^\circ$\\
CSS & $3.81^\circ$ & $5.58^\circ$ & $5.52^\circ$ & $7.34^\circ$ \\
Proposed & $1.41^\circ$ & $1.62^\circ$ & $1.68^\circ$ & $1.75^\circ$ \\
\bottomrule
 \end{tabular}
 \caption{Single white noise source}
 \label{table:alterwhite10db}
\end{table}

With the same background noise conditions, the second experiment features two white noise sources being played by both speakers in a competing fashion. The result is shown in Table~\ref{table:twowhite10db}. The MAE of hist-ESPRIT is slightly lower than the proposed algorithm as the error influence of estimation in low frequencies is slightly lower, while the SDE for the proposed algorithm is clearly superior.
\begin{table}
\centering
\begin{tabular}{ccc|cc}
 \toprule
 & \multicolumn{2}{c|}{SNR=10dB}& \multicolumn{2}{c}{SNR=0dB} \\
 Algorithm & MAE & SDE & MAE & SDE \\
 \hline
 \midrule
hist-ESPRIT & $2.57^\circ$ & $5.64^\circ$ & $3.28^\circ$ & $7.47^\circ$\\
CSS & $6.8^\circ$ & $9.67^\circ$ & $6.5^\circ$ & $12.22^\circ$ \\
Proposed & $2.82^\circ$ & $3.55^\circ$ & $2.75^\circ$ & $5.62^\circ$ \\
\bottomrule
 \end{tabular}
%
%
 \caption{Two simultaneous white noise sources}
 \label{table:twowhite10db}
\end{table}

The final experiment used the same scenario as the second experiment, but two simultaneously active speech sources were to be localized. The level of the speech signals changed over time, and the estimated SNR was in the range of $[-5,10]$dB. The result is shown in Table~\ref{table:twospeech10db}. The performance of the proposed algorithm is better than CSS algorithm, but worse than the hist-ESPRIT algorithm, especially on SDE. The speech signal is spectrally sparse and the energy  is compacted in the fundamental frequency and its harmonics.
In contrast, under the white noise background, the intervals between harmonics in the spectrum are noisier (lower SNR). With the inspection of the narrowband processing, the estimated DOA results from the intervals had large errors (up to $90^\circ$). The reason behind the reduction in performance of the proposed algorithm is the inferior robustness of the least squares criterion in comparison to the histogram method. In the white sources experiments above, the SNR is constant for the entire frequency range. Therefore, those experiments had better results.

\begin{table}
\centering
\begin{tabular}{cccc}
 \toprule
 Algorithm & MAE & SDE \\
 \hline
 \midrule
hist-ESPRIT & $4.28^\circ$ & $6.47^\circ$\\
CSS & $6.85^\circ$ & $15.37^\circ$ \\
Proposed & $5.75^\circ$ & $10.62^\circ$ \\
\bottomrule
 \end{tabular}
 \caption{Two simultaneous talkers, SNR$\in [-5,10]$dB}
 \label{table:twospeech10db}
\end{table}

Through the above experiments, compared to the computing time of hist-ESPRIT algorithm, CSS algorithm (with $0.1^\circ$ resolution of the spatial spectrum) was $13\%$ faster, and the proposed algorithm was $22.6\%$ faster.

\section{Conclusion}
\label{sec:conclusion}

A wideband signal subspace DOA estimation approach is presented. The proposed signal subspace rotation method and the narrowband signal covariance matrix reconstruction method are high and outperform the existing conventional approaches in computational complexity. Additionally, the proposed approach avoids the necessity of additional prior knowledge for extending the narrowband ESPRIT to a wideband scheme. Experiments based on real recordings validate the effectiveness and low computational complexity of the proposed method.
As part of the future work, the proposed algorithm should be analyzed in the context of solving the spatial aliasing problem, in order to improve the localization accuracy of the spectrally sparse sources in environments with background noise.

\bibliographystyle{IEEEbib}
\bibliography{Bibliography}

\begin{thebibliography}{10}

\bibitem{jin17}
W.~{Jin}, M.~J. {Taghizadeh}, K.~{Chen}, and W.~{Xiao},
\newblock ``Multi-channel noise reduction for hands-free voice communication on
  mobile phones,''
\newblock in {\em 2017 IEEE International Conference on Acoustics, Speech and
  Signal Processing (ICASSP)}, March 2017, pp. 506--510.

\bibitem{jin18}
W.~{Jin}, B.~{Desikan}, A.~{Kumar}, and K.~{Chen},
\newblock ``Multi-channel noise reduction with interference suppression on
  mobile phones,''
\newblock in {\em 2018 16th International Workshop on Acoustic Signal
  Enhancement (IWAENC)}, Sep. 2018, pp. 201--205.

\bibitem{phat}
C.~Knapp and G.~Carter,
\newblock ``The generalized correlation method for estimation of time delay,''
\newblock {\em Acoustics, Speech and Signal Processing, IEEE Transactions},
  vol. 24(4), pp. 320 -- 327, 1976.

\bibitem{srpphat}
J.~H. DiBiase,
\newblock {\em A High-Accuracy, Low-Latency Technique for Talker Localization
  in Reverberant Environments Using Microphone Arrays},
\newblock Ph.D. thesis, Brown University, 2000.

\bibitem{music}
R.~Schmidt,
\newblock ``Multiple emitter location and signal parameter estimation,''
\newblock {\em Antennas and Propagation, IEEE Transactions}, vol. 34(3), pp.
  276 -- 280, 1986.

\bibitem{esprit}
R.~Roy and T.~Kailath,
\newblock ``{ESPRIT}-estimation of signal parameters via rotational invariance
  techniques,''
\newblock {\em Acoustics, Speech and Signal Processing, IEEE Transactions},
  vol. 37(7), pp. 984--995, 1989.

\bibitem{lombardtrans}
A.~Lombard, Y.~Zheng, H.~Buchner, and W.~Kellermann,
\newblock ``{TDOA} estimation for multiple sound sources in noisy and
  reverberant environments using broadband independent component analysis,''
\newblock {\em IEEE Transactions on Audio, Speech, and Language Processing},
  vol. 19(6), pp. 1490 -- 1503, 2011.

\bibitem{nesta2011}
F.~Nesta, P.~Svaizer, and M.~Omologo,
\newblock ``Convolutive {BSS} of short mixtures by {ICA} recursively
  regularized across frequencies,''
\newblock {\em IEEE transactions on audio, speech, and language processing},
  vol. 19, no. 3, pp. 624 -- 639, 2011.

\bibitem{andreas2018cluster}
A.~Brendel, C.~Huang, and W.~Kellermann,
\newblock ``{STFT} bin selection for localization algorithms based on the
  sparsity of speech signal spectra,''
\newblock in {\em European Congress and Exposition on Noise Control
  Engineering}. IEEE, 2018, pp. 2561--2568.

\bibitem{araki2006doa}
S.~Araki, H.~Sawada, R.~Mukai, and S.~Makino,
\newblock ``{DOA} estimation for multiple sparse sources with normalized
  observation vector clustering,''
\newblock in {\em IEEE International Conference on Acoustics, Speech, and
  Signal Processing (ICASSP)}. IEEE, 2006, vol.~5.

\bibitem{eb_esprit}
H.~Teutsch and W.~Kellermann,
\newblock ``{EB-ESPRIT}: 2{D} localization of multiple wideband acoustic
  sources using eigenbeams,''
\newblock {\em IEEE International Conference}, vol. iii/89 - iii/92 Vol. 3(3),
  pp. 89--92, 2005.

\bibitem{Khaykin2009Coherent}
D.~Khaykin and B.~Rafaely,
\newblock ``Coherent signals direction-of-arrival estimation using a spherical
  microphone array: Frequency smoothing approach,''
\newblock in {\em Applications of Signal Processing to Audio and Acoustics,
  2009. WASPAA '09. IEEE Workshop on}, 2009, pp. 221--224.

\bibitem{css}
H.~Wang and M.~Kaveh,
\newblock ``Coherent signal-subspace processing for the detection and
  estimation of angles of arrival of multiple wide-band sources,''
\newblock {\em IEEE Transactions on Acoustics, Speech, and Signal Processing},
  1985.

\bibitem{cssesprit1}
A.~Shaw and R.~Kumaresan,
\newblock ``Estimation of angles of arrivals of broadband signals,''
\newblock in {\em IEEE International Conference on Acoustics, Speech, and
  Signal Processing (ICASSP)}, 1987.

\bibitem{cssesprit2}
Y.-H. Chen and R.-H. Chen,
\newblock ``Directions-of-arrival estimations of multiple coherent broadband
  signals,''
\newblock {\em IEEE transactions on aerospace and electronic systems}, vol. 29,
  no. 3, pp. 1035 -- 1043, 1993.

\bibitem{focusingmatrix}
H.~Hung and M.~Kaveh,
\newblock ``Focussing matrices for coherent signal-subspace processing,''
\newblock {\em IEEE Transactions on Acoustics, Speech, and Signal Processing},
  vol. 36, no. 8, pp. 1272 -- 1281, 1988.

\bibitem{espritextension}
B.~Ottersten and T.~Kailath,
\newblock ``Direction-of-arrival estimation for wide-band signals using the
  {ESPRIT} algorithm,''
\newblock {\em IEEE Transactions on Acoustics, Speech, and Signal Processing},
  vol. 38, no. 2, pp. 317 -- 327, 1990.

\bibitem{tensoresprit}
F.~Raimondi, P.~Comon, and O.~Michel,
\newblock ``Wideband multilinear array processing through tensor
  decomposition,''
\newblock in {\em IEEE International Conference on Acoustics, Speech and Signal
  Processing (ICASSP)}, 2016.

\bibitem{broadbeam}
D.~B. Ward, Z.~Ding, and R.~A. Kennedy,
\newblock ``Broadband {DOA} estimation using frequency invariant beamforming,''
\newblock {\em IEEE Transactions on Signal Processing}, 1998.

\bibitem{replicationphase}
K.~Chen, J.~T. Geiger, W.~Jin, M.~Taghizadeh, and W.~Kellermann,
\newblock ``Robust phase replication method for spatial aliasing problem in
  multiple sound sources localization,''
\newblock in {\em IEEE Workshop on Applications of Signal Processing to Audio
  and Acoustics (WASPAA)}, 2017.

\bibitem{esprithistogram}
R.~Roy, A.~Paulraj, and T.~Kailath,
\newblock ``Comparative performance of {ESPRIT} and {MUSIC} for
  direction-of-arrival estimation,''
\newblock in {\em IEEE International Conference on Acoustics, Speech, and
  Signal Processing (ICASSP)}, 1987.

\bibitem{gridcopus}
M.~Cooke, J.~Barker, S.~Cunningham, and X.~Shao,
\newblock ``An audio-visual corpus for speech perception and automatic speech
  recognition,''
\newblock {\em The Journal of the Acoustical Society of America}, vol. 120(5),
  pp. 2421 -- 2424, 2006.

\end{thebibliography}

\end{document}